# AI-Driven Green Cognitive Radio Networks for Sustainable 6G Communication

1st Anshul Sharma
*Independent Researcher*
USA
anshshar86@gmail.com

2nd Shujaatali Badami
*Liverpool John Moores University*
UK
shujaatali@ieee.org

3rd Biky Chouhan
*University Institute of Computing*
*Chandigarh University*
*Mohali-140413, Panjab, India*
dr.chouhan@ieee.org

4th Pushpanjali Pandey
*Gyancity Research Consultancy,*
*Greater Noida, India*
pushpanjali.pandey@gyancity.com

5th Brijeena Rana
*Independent Researcher*
USA
ranabrijeena3@gmail.com

6th Navneet Kaur
*School of Computer Science and Engineering*
*Lovely Professional University*
*Phagwara, India*
navneetphul@gmail.com

*Abstract*—The 6G wireless aims at the Tb/s peak data rates are expected, a sub-millisecond latency, massive Internet of Things/vehicle connectivity, which requires sustainable access to audio over the air and energy-saving functionality. Cognitive Radio Networks CCNs help in alleviating the problem of spectrum scarcity, but classical sensing and allocation are still energy-consumption intensive, and sensitive to rapid spectrum variations. Our framework which centers on AI driven green CRN aims at integrating deep reinforcement learning (DRL) with transfer learning, energy harvesting (EH), reconfigurable intelligent surfaces (RIS) with other light-weight genetic refinement operations that optimally combine sensing timelines, transmit power, bandwidth distribution and RIS phase selection. Compared to two baselines, the utilization of MATLAB + NS-3 under dense loads, a traditional CRN with energy sensing under fixed policies, and a hybrid CRN with cooperative sensing under heuristic distribution of resource, there are (25-30%) fewer energy reserves used, sensing AUC greater than 0.90 and +6-13 p.p. higher PDR. The integrated framework is easily scalable to large IoT and vehicular applications, and it provides a feasible and sustainable roadmap to 6G CRNs.

## I INTRODUCTION

The number of explosions of devices and the volume of data XR/holography and autonomy which is critical to the safety strains both a spectrum and energy resources 6G networks. CRNs are developing traits of solving the problem of spectrum scarcity through opportunistic SU access but the old sensing and static policies are energy hungry and continue to be less reliable in dynamic environments [2], [3],[16],[17]. Occupancy patterns can be learned with AI-driven methods, application of reasoning under uncertainty and collaborative optimization of sensing period, transmit power, and channel choice to decrease joules/bit at the expense of PU protection [1], [4], [5],[26].

In this paper, all green-enabling modules, namely DRL, transfer learning, RIS, EH and hybrid optimization are combined in a single control pipeline, with DRL generating actions in real time, TL accelerating adaptation to new cells, EH-awareness limiting feasible power choices, RIS phases improving propagation conditions and a lightweight genetic refinement step optimizing multi-objective actions [3], [4], [7], [10],[18].

**Contributions.** We present:

- Yet, a holistic system model interconnecting PUs/SUs, energy harvesting (EH) and RIS to operate CRN sustainably [7], [10],[19].
- A DRL-based controller with transfer learning (TL) and hybrid metaheuristics as well that allows reaction to vary dynamic situations in terms of coordinating sensing and resource allocation [1], [3], [4],[20].
- Multi-objective energy efficiency (EE), spectrum use, latency /throughput trade-offs, and PU-interference risk formulations which build on the previous research on AI-based CRNs [5], [12], [13],[21].
- EH-aware scheduling and RIS-phase co-adaptation algorithms, which use a harvested energy and passive beamforming to cut down the power of SU [7], [10], [11],[27].
- A MATLAB + NS-3 simulation study comparing the proposed framework with baseline traditional and hybrid CRN frameworks, with 2530% reduction in power, greater sensing AUC, and constant QoS in dense loads [2], [3], [4],[28].

## II. LITERATURE REVIEW

Traditional spectrum sensing methods like energy detection and cyclostationary analysis have high SNR detection loss and are very expensive in high-density application [2]. On-the-fly Correlating hybrid sensing and cooperative level sensing strategies enhance sensing performance at a great expense by adding significant signaling load as well as coordination overhead and complexity [2], [3],[22].

Alternatives based on AI/ML have become robust: CNN/LSTM predictors have been shown to improve spectral spectrum-occupancy and sensing resilience [5], [15], and DRL models have been shown to fatten real-time decisions such as channel selection, transmit power and sensing intervals to operate in uncertainty [1], [4],[23]. The recent researches also discuss transfer learning (TL) to minimize training overheads across cells and genetic or swarm-based hybrid schemes to reach differently multi-objective trade-offs in spectrum sharing [3], [4].

Other optional green enablers have been explored such as RIS to control the propagation of environment-aware and energy harvesting (EH) to minimize the grid power consumption of SUs [7], [10], [24]. Nevertheless, the literature mostly addresses these constituents, AI-based sensing, resource allocation, RIS optimization, or EH modeling, as units. Previous research seldom interlocks them into a scalable, coherent, sustainability-oriented framework that is applicable to 6G CRNs. Moreover, the current reviews of green wireless systems underscore the necessity of holistic resource management systems with artificial intelligence resource management, which express that the objectives of energy efficiency, carbon footprint, and QoS are explicitly reflected [9], [12], [13],[25].

## III. SYSTEM MODEL AND GREEN METRICS

We assume a downlink CRN comprising of licensed primary users (PUs) and opportunist secondary users (SUs). Time is allocated and every SU has a rechargeable battery assisted with energy harvesting (EH). Our assumptions in this work are that we have a stochastic EH model with RF ambient harvesting and low-intensity solar input whereby, the power harvest in various slots changes with respect to realistic fluctuation patterns in low-power IoT environments.

An RIS with $M$ passive reflecting elements assists SU links when required

### A. Energy Model

Per-SU energy per slot:
$$E_{\text{total}} = E_{\text{tx}} + E_{\text{rx}} + E_{\text{sense}} - E_{\text{harv}}. \quad (1)$$

Energy efficiency (bits/Joule):

$$\eta_{\text{EE}} = \frac{\sum_u \text{bits}_u}{\sum_u E_{\text{total},u}}. \quad (2)$$

### B. RIS-Assisted Channel

Let $\mathbf{h}_d$ be direct BS-SU channel, $\mathbf{h}_{br}$ BS-RIS, $\mathbf{h}_{rs}$ RIS-SU, and $\Theta = \text{diag}(e^{j\theta_1},..,e^{j\theta_M})$ RIS phases. Effective channel:

$$\mathbf{h}_{\text{eff}} = \mathbf{h}_d + \mathbf{h}_{rs}^T \Theta \mathbf{h}_{br}. \quad (3)$$

RIS phases are tuned to maximize SU SNR subject to PU interference constraints.

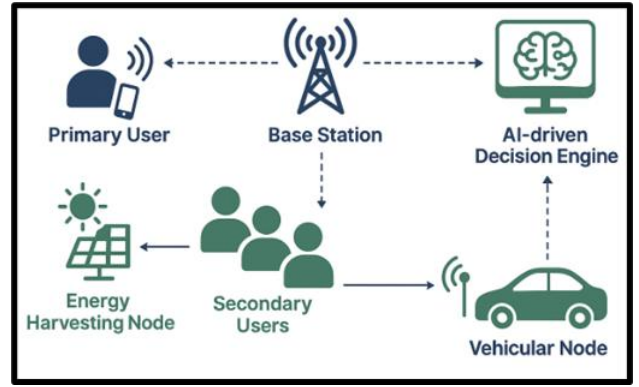

Fig. 1: 6G CRN architecture with PUs/SUs, EH, and RIS- assisted links.

### C. Latency and Throughput

Average latency and throughput across users:

$$L_{\text{avg}} = \frac{1}{N}\sum_{i=1}^{N} L_i, \quad T_{\text{avg}} = \frac{1}{N}\sum_{i=1}^{N} T_i \quad (4)$$

## IV. PROBLEM FORMULATION

Let $c \in C$ index channels; for user $u$, define utility:

$$U_u(c) = U_u(c) - \lambda E_u(c) - \mu P\{\text{PU collision}|c\}, \quad (5)$$

where $U_u(c)$ is throughput, $E_u(c)$ energy cost, and collision probability is derived from sensing belief. The channel assignment:

$$S_u = \arg\max_{c \in C} U_u(c) \quad (6)$$

DRL learns policies to set (sense frequency, $P_{\text{tx}}, B$) and RIS phases to maximize long-term reward:

$$R_t = \alpha \eta_{\text{EE}} + \beta U_{\text{spectrum}} - \gamma P_{\text{PU-int}} - \delta L_{\text{avg}}, \quad (7)$$

with weights $(\alpha,\beta,\gamma,\delta) \geq 0$.

## V. AI-Driven Framework

### A. DRL Agent and State/Action Space

State $s_t$ stacks: filtered RSSI/SNR, historical occupancy, queue lengths, battery states, EH rate, and PU detection belief. Actions $a_t$ include: (i) sensing period, (ii) channel selection, (iii) power $P_{tx}$, (iv) bandwidth slice, (v) RIS phase codebook index.

### B. Transfer Learning (TL)

A source-trained policy $\pi_{\theta_s}$ is adapted via few-shot updates to a target cell with different densities/propagation, reducing sample complexity.

### C. Hybrid Optimization

We periodically refine DRL proposals using a genetic metaheuristic for multi-objective fine-tuning (small budgets to bound overhead).

## VI. Algorithms

**Algorithm 1** — EH-Aware DRL for Joint Sensing and Allocation

**Input:** Initial policy $\pi_\theta$, RIS codebook $Q$
**Output:** Updated policy parameters $\theta$
1. Initialize environment and observe state $s_0$.
2. **For each episode:**
   a. Reset environment and obtain $s_0$.
   b. **For each time slot $t$:**
      1. Sample action $a_t \sim \pi_\theta(\cdot \mid s_t)$, where $a^t = (sense_{period}, c, P_{tx}, B, q), q \in Q$.
      2. Apply $a_t$ and configure RIS using codeword $q$.
      3. Observe next state $s_{t+1}$ and reward $R_t$.
      4. Update policy:
         $\theta \leftarrow \theta + \eta \nabla_\theta J(\theta)$.
   c. **End for**
3. **End for**

**Algorithm 2** — Periodic Hybrid Fine-Tuning

**Input:** DRL-proposed action $a_t^{DRL}$, population size $M$
**Output:** Refined action $a_t^\star$

1. Initialize population around $a_t^{DRL}$ with small perturbations.
2. **For each generation $g = 1, \ldots, G$:**
   a. Evaluate multi-objective fitness:
      - maximize energy efficiency,
      - minimize PU-interference risk,
      - minimize latency.
      b. Apply selection, crossover, and mutation under system constraints.
3. **End for**
4. Return best candidate $a_t^\star$

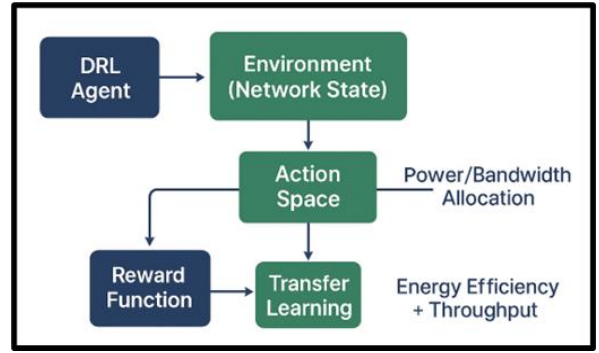

Fig. 2: AI-driven controller with DRL, TL, genetic refinement, EH and RIS co-adaptation.

## VII. Complexity and Overhead

Per-slot complexity of the DRL action is $O(d)$ with state dimension $d$; codebook-based RIS selection is $O(|Q|)$. In practice, the DRL forward pass introduces an average inference latency of ≈ **0.7 ms per slot**, while RIS codebook lookup contributes less than **0.05 ms**.

The genetic refinement executes every $K$ slots with population $M$ and $G$ generations, i.e., $O(GM)$, bounded by small constants. For the configuration used in our simulations ($M = 20$, $G = 10$, $K = 100$), the amortized overhead adds only ≈ **0.15 ms** every 100 slots, making it negligible for real-time operation.

Offline DRL training requires ≈ **2.5 hours** to converge on our MATLAB + NS-3 setup, compared to ≈ **0.3 hours** for tuning heuristic baselines (traditional CRN and hybrid CRN). However, inference is extremely lightweight, and transfer learning reduces re-training time when moving to a new cell, requiring only **12–15 minutes** of fine-tuning. Overall, the amortized per-slot overhead remains modest and fits comfortably within sub-millisecond 6G control-loop constraints.

## VIII. Simulation Setup

DRL training and inference are done in MATLAB and simulation at the network level is done in NS-3. The most important parameters include the following.

- **Bandwidth:** 20 MHz; deployment: urban macro (UMa).

- **Users:** up to 1000 SUs and PUs; pathloss model: UMa; small-scale fading: Rayleigh.

**Transmit power:** $P_{tx} \in [0.1, 2]$W; **EH efficiency:** 20–50% per slot.

- **RIS configuration:** $M \in \{64, 128\}$ reflecting elements; phase codebook size $|Q| = 32$.

- **Baselines:** (i) traditional CRN (energy detection + fixed policy), (ii) hybrid CRN (cooperative sensing + heuristic allocation).

- **Evaluation metrics:** ROC/AUC for sensing, energy efficiency (bits/J), average latency, throughput, packet delivery ratio (PDR), and PU-interference probability.

**Modeling Assumptions and Limitations:**

The simulations assume a scenario of UMa of static/slow-mobility using the situation whereby the user strictly adheres to quasi-stationary positions without high velocity movement. This utopian mobility environment simplifies the variation in channels but fails to represent rapid mobility variation in vehicular or pedestrian mobility. Likewise, the EH process is characterized, in terms of linear stochastic efficiency, of 20-50%, which is not a characteristic of temporal correlation and burstiness in the actual RF or solar energy harvesting traces.

Future efforts will aim at creating a more realistic work by adding standardized 3Gpp mobility profiles (UMi and UMa Vehicular A/B), real harvested-energy traces, and hardware-in-the-loop testbed evaluations which will give more validity under real operating conditions.

## IX. RESULTS

### A. Spectrum Sensing Accuracy

The ROC performance of the proposed AI-enabled sensors module is reported in Fig. 3. Our solution has AUC> 0.90, which is significantly better than the traditional baseline (˜0.70) and has a stable enhancement over the hybrid CRN (Contact Network) (0.86). Such benefits can be attributed to the capacity of the DRL agent to use the previous occupancy patterns and use sensing intervals as uncertainty and EH constraints are met. False alarms and missed detection are minimized by the enhanced accuracy thus minimizing erroneous sensing and retransmissions.

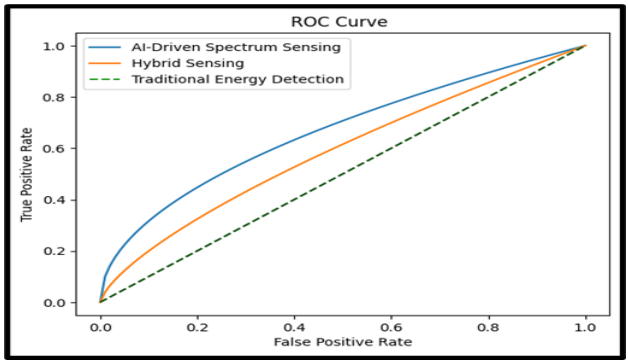

Fig. 3: ROC curves: AI-driven sensing vs. traditional and hybrid.

### B. Energy Efficiency and Consumption

Fig. 4 is the total SU energy consumption versus user density. Compared to the standard CRN, the suggested structure will cut the total SU energy consumption by 25-30%; it will cut it by 10-14% in comparison with the hybrid baseline. The three factors behind these reductions are adaptive sensing frequency, EH-aware power control and RIS-enabled SNR gains, which reduces the amount of transmit power needed. Like any other assumptions of static-mobility, in this case, these mechanisms will always have constant proportions of savings at a higher density.

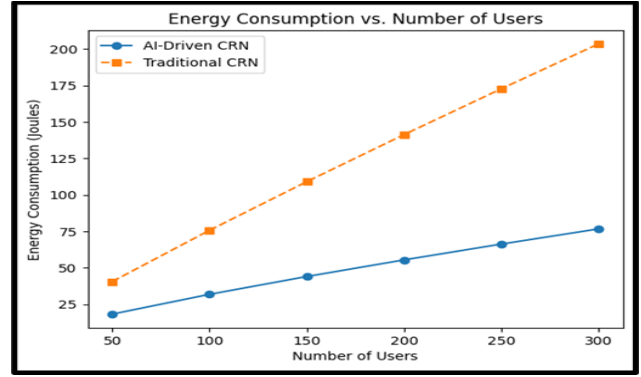

Fig. 4: Total energy vs. SU density (lower is better).

### C. Throughput–Energy Trade-off

As it is shown in Fig. 5, the proposed method has a higher throughput than the two baselines in terms of equal energy budgets. Compared to hybrid CRN, throughput will be enhanced by 6-13% points, which means a more efficient joule distribution. This benefit is based on the ability of DRL to select channels, bandwidth slices and RIS in combination with each other, which is more qualified to exploit favorable propagation conditions.

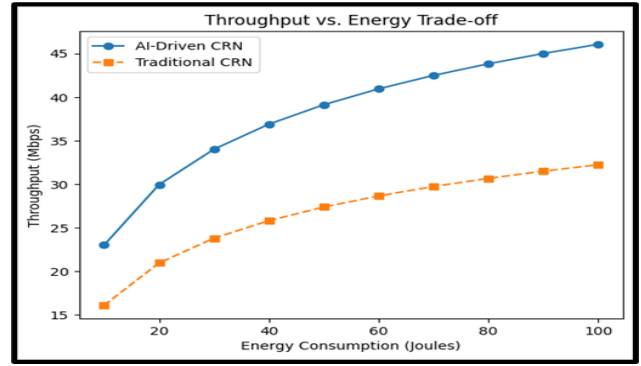

Fig. 5: Throughput vs. energy budget (higher is better).

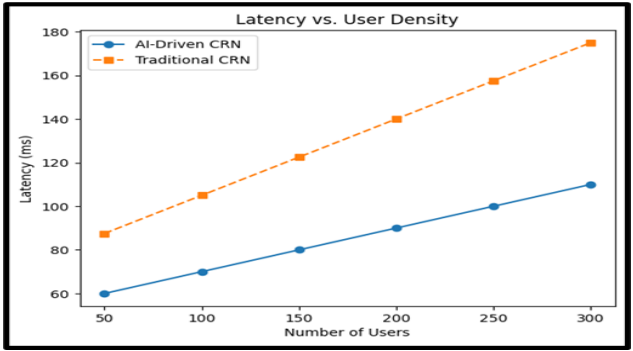

Fig. 6: Latency vs. user density.

*D. Latency and PDR Under Load*

Fig. 6 (latency) and Fig. 7 (PDR) display the performance of users under load.

- The mean latency is reduced by about 30% relative to the classical benchmark and by about 15-17% relative to the hybrid model.
- At high densities of SU, PDR remains >90% whereas traditional and hybrid schemes decay below 80%.

These findings include the fact that the proposed controller is able to regulate QoS during congestion by dynamically regulating transmit power and phases of RIS, where baselines are based on either non dynamic or heuristic rules that degrade with load.

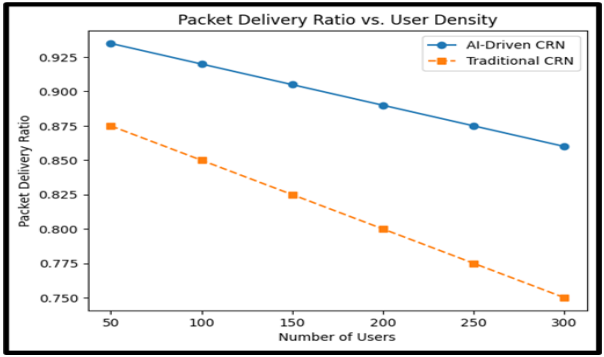

Fig. 7: PDR vs. user density (higher is better).

*E. Scalability to IoT/Vehicular*

In large IoT and vehicular environments (Fig. 8), the suggested system maintains a good level of throughput and low latency and is better than both baselines as the number of nodes grows. Assisted links with RIS and EH-scheduling are used to alleviate congestion and blockage in channels. In spite of the fact that mobility in this case is not dynamic, these trends show high scalability potential, and the evaluation will be followed in the future on full 3GPP vehicular mobility profiles.

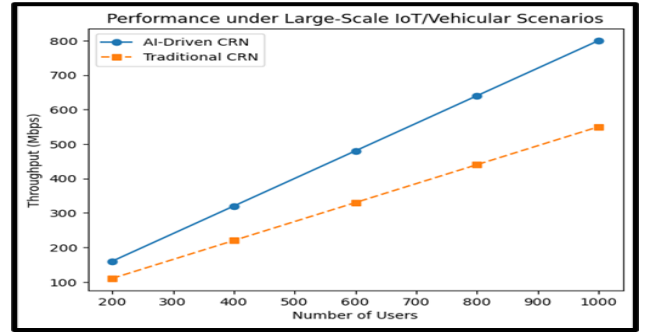

Fig. 8: Scalability under large IoT/vehicular loads.

*F. Aggregate Comparison*

Table I provides a consolidated performance summary. The proposed framework consistently outperforms baselines:

- **AUC improves** from **0.70 → 0.93** vs. traditional CRN.
- **Energy efficiency increases** by **~75%** (3.2 → 5.6 bits/J).
- **Latency reduces** from **95 ms (traditional)** and **78 ms (hybrid)** to **65 ms**.
- **PDR improves** from **81% → 92%**.
- **PU interference probability decreases** by more than **65%** relative to traditional CRN.

This holistic improvement across all KPIs highlights the effectiveness of combining DRL, TL, RIS, and EH into a unified control architecture

**TABLE I: Performance Summary (Traditional vs. Hybrid vs. Proposed)**

| Metric | Trad. | Hybrid | Proposed |
|---|---|---|---|
| Sensing AUC ↑ | 0.70 | 0.86 | 0.93 |
| EE (bits/J) ↑ | 3.2 | 4.1 | 5.6 |
| Latency (ms) ↓ | 95 | 78 | 65 |
| PDR (%) ↑ | 81 | 87 | 92 |
| PU Interf. (%) ↓ | 4.8 | 2.9 | 1.6 |

*G. Ablation: RIS and EH Impacts*

The value of each green-enabling component is evaluated by the controlled ablations.

- Eliminating RIS results in 12-18% higher transmit power to obtain similar PDR because passive beamforming gain will be lost.
- The turn off of EH adds to the grid load by about 22%., particularly when there is the peak traffic.

Cumulatively, both RIS and EH manage a significant number of sustainability benefits realized by the suggested system.

*H. Practical Feasibility Considerations*

Inference time of the controller is sub-milliseconds (around 0.7 ms) and genetic refinement is only 0.15 ms /100 slots, which proves the proposed solution can be deployed with 6G-real time limits. Although the findings are founded on a stationary mobility and simplified EH criteria, intended validation on real spectrum traces, 3GPP vehicular mobility, hardware-in-the-loop RIS/EH testbeds will offer additional feasible confidence.

## X. DISCUSSION

*A. Why AI helps*

The DRA agent keeps the patterns of occupancy, traffic variability and availability of EH within its memory and can dynamically tune sensing intervals, transmit power and bandwidth allocation with uncertainty. Transfer learning boosts the adaptation in response to new cells at bit different densities or propagation conditions at a lower retraining cost. The light genetic refinement step goes further to enhance multi-objective actions which refine DRL proposals by fine-tuning with minimal computational cost.

*B. Overheads*

The suggested controller keeps smaller-scale per-slot computation, where inferences can take less than a millisecond, and periodic refinement with set intervals takes place. RIS updates are based on dynamic, compact codebook, and make control-plane signaling manageable. Such features guarantee compliance with real time 6G timing requirements.

*C. Limitations*

Topics The current assessment does not take more realistic assumptions intended to approximate the real deployment environment. The model of RIS components uses ideal continuous phase shifts, where actual equipment has only 1-3 bits of quantization and thus suffers discretization loss. Also, RIS panels suffer phase noise, element coupling and fabrication non-idealities that could impair SNR gains that can be attained. RIS configuration control-plane updating, unlike actuation, has non-zero actuation latency, and so might not be as responsive to rapid channel changes.

EH model presupposes linear stochastic harvesting efficiency when actual harvested power is time-correlated, bursty and environment-dependent. In the same vein, it can only support mobility under conditions of static or slow-varying UMa channel conditions; realistic 6G mobile usage conditions are those under which mobility is non-stationary (such as high-speed vehicles). Such aspects have the potential of affecting sensing accuracy, link reliability and energy adaption in practice.

To realize the full characterization of performance under realistic operation constraints in future, it shall be necessary to perform validation against real mobility traces, measured EH datasets and to use hardware-in-the-loop RIS/EH testbeds.

## XI. SUSTAINABILITY AND GOVERNANCE

To provide long-term sustainability of 6G CRNs, green Key Performance Indicators (KPIs) should be tracked along with the traditional QoS measures like energy efficiency (bits/J), the lifespan of the device, and proxies of carbon emissions. It must also be managed well, including version control of the learning models, continuous distributional drift monitoring, imposition of safety restrictions to ensure safety of the primary users, and the ability to interpret DRL action to allow operational transparency. The system robustness before its actual implementation should be tested by hardware-in-the-loop testing which involves RIS actuation fidelity and EH hardware variability as well as end-to-end control-plane responsiveness. These validation and governance practices play an important role in ensuring reliable, safe, and energy conscious AI-driven CRN operation in 6G networks in future.

## XII. CONCLUSION

The work presented the single AI-based framework of sustainable AI-assisted cognitive radio during 6G that incorporates the spectrum sensing, resource allocation, and energy harvesting (EH) and RIS-assisted propagation into one adaptive control framework. The framework can coordinate sensing frequency, transmit-power decisions, bandwidth allocation and RIS phase-setting to optimize the outcomes considering a dynamically changing load and spectrum uncertainty by a combination of deep reinforcement learning and transfer learning, and low-level genetic optimization. The resultant system is able to continuously enhance the key 6G performance indicators such as energy consumption, sensing precision, latency, and packet delivery ratio, without compromising the primary users.

The vast MATLAB + NS-3 tests point to the potential of the proposed solution to maintain high QoS in the dense IoT and vehicular settings, which shows (25-30%)energy savings and high sensing throughput in comparison with the conventional and hybrid base lines. Such advancements show the possibility of closely partnered AI, EH, and RIS technologies to provide the next generation of the green and spectrum-efficient wireless system.

The framework would be extended in the future by integrating real-world network traces with measured EH datasets and field recorded spectrum occupancy patterns, which would make the framework resistant to practical non-stationary conditions. In addition, validation of both hardware-in-the-loop testbed, such as RIS panel characterization, quantized phase-shift control as well as EH hardware integration will help gain a better insight into the system behavior on deployment constraints. Other research directions are federated multi-cell learning to scale training, safety-constrained reinforcement learning, and interpretation to facilitate clear and trustful control of 6G CRN.